\documentclass[journal]{IEEEtran}

\ifCLASSINFOpdf
\else
	\usepackage[dvips]{graphicx}
\fi
\usepackage{url}
\usepackage{graphicx}
\usepackage{subfig}
\usepackage{multirow}
\usepackage{footnote}
\usepackage{bm}
\usepackage{epstopdf}
\usepackage{amsmath}
\usepackage{amssymb}
\usepackage{multicol,lipsum}
\usepackage{color,xcolor}
\usepackage{amsthm}
\usepackage{cite}
\usepackage{algorithm}
\usepackage{algpseudocode}
\usepackage{threeparttable}
\usepackage{cuted}
\usepackage{mathrsfs}

\begin{document}

\title{One-bit Spectrum Sensing with the Eigenvalue Moment Ratio Approach}

\author{Yuan Zhao~\IEEEmembership{Member,~IEEE}, 
	Xiaochuan Ke, Bo Zhao, Yuhang Xiao,
	Lei Huang~\IEEEmembership{Senior Member,~IEEE} 
	\thanks{The authors are with the School of Electrical and Information Engineering, Shenzhen University, Guangdong, China}.
	\thanks{Manuscript received XXX, XX, 2020; revised XXX, XX, 2020. The work described in this letter was supported by the National Science Fund for Distinguished Young Scholars under Grant 61925108 and National Natural Science Foundation of China under Grant U1713217. ($Corresponding\ author: Bo\ Zhao $)}
}

\markboth{IEEE WIRELESS COMMUNICATIONS LETTERS,~Vol.~14, No.~8, August~2021}
{Shell \MakeLowercase{\textit{et al.}}: Bare Demo of IEEEtran.cls for IEEE Journals}
%

\maketitle

\begin{abstract}
	One-bit analog-to-digital converter (ADC), performing signal sampling as an extreme simple comparator, is an overwhelming technology for spectrum sensing due to its low-cost, low-power consumptions and high sampling rate. In this letter, we propose a novel one-bit sensing approach based on the eigenvalue moment ratio (EMR), which has been proved to be highly efficient for conventional multi-antenna spectrum sensing in $\infty$-bit situation. Particularly, we determine the asymptotic distribution of one-bit EMR under null hypothesis via the central limited theorem (CLT), allowing us to perform spectrum sensing with one-bit samples directly. Theoretical and simulation analysis show the new approach can provide reasonably good sensing performance at a low hardware cost.
\end{abstract}

\begin{IEEEkeywords}
	Spectrum sensing, one-bit quantization, signal detection, central limited theorem.
\end{IEEEkeywords}

\IEEEpeerreviewmaketitle

\section{Introduction}
\IEEEPARstart{C}{ognitive} radio (CR) is a promising technique for improving the efficiency of spectrum resource allocations \cite{RN7}. One of the fundamental applications of CR is spectrum sensing, which allows the secondary users (SUs) to occupy the frequency band when the primary users (PUs) are not active \cite{RN1, RN8, RN9}. This requires the SU to frequently listen to the full frequency band, in other words, detect the presence of the PUs, in order to vacate the channel within the required time delay. However, with the development of next generation wireless communication systems, the bandwidth to observe has been substantially expanded, posting an increasing demand on the sampling rate of the analog-to-digital convertors (ADCs). As there is a trade-off between sampling rate and quantization precision, under the same hardware and power budget, the usage of low resolution quantizers, particularly the one-bit ADCs, is rather beneficial in monitoring the ultra-wide spectrum bands.      

In the meantime, recent developments on the one-bit signal processing have attracted attentions in wireless communication community including \cite{RN10, RN11} and the references therein. However, only a few works address the spectrum sensing with one-bit samples, e.g. \cite{RN4, RN6}. Still, this issue has not been investigated for the multi-antenna scenario. In fact, one-bit spectrum sensing is practically challenging from the following two aspects. Firstly, the amplitude of the raw data is discarded due to the extreme quantization scheme, thus leading to the failure of the traditional energy based detectors. On the other hand, the likelihood function for the output of the one-bit quantizer involves the orthant probability \cite{RN2}, which remains an open problem for dimensions greater than 3. This prohibits the accurate formulation of closed-from testing statistics following standard criteria such as the generalized likelihood ratio. Hence, the existing literatures either require the \emph{a priori} information on covariance matrix \cite{RN12} or assume the independent of receiving channel \cite{RN6, RN13, RN14}, which is somehow not realistic for practical systems.

Fortunately, the well-established \emph{arcsin} law formulates the connection between the second order statistics before and after the one-bit quantization \cite{RN15}. It is worth pointing out that the one-bit quantization inherits the correlation structure of the conventional quantization schemes, which equivalently casts the detection problem into a test for the sphericity of the received signal. As a popular spectrum sensing approach, the eigenvalue moment ratio (EMR) \cite{RN1} algorithm test sphericity by computing the ratio between the $r$th moment of the sample eigenvalues and the $r$th power of the their summation. Moreover, the distribution of EMR statistic is accurately determined under generalized asymptotic regime ($m,n \to \infty$ and $m/n \to c \in \left(0, \infty \right)$). Nevertheless, this distribution is obtained under the Gaussian assumption, by which one-bit samples do not satisfy.

Therefore, in this letter, we focus on the statistically analysis on the one-bit EMR algorithm with $r = 2$ (coincidentally identical to the John’s detector \cite{RN16, RN17}). We chose to analyze its distribution in the asymptotic regime, where the central limit theorem (CLT) is applied to convert the discrete probabilities into continuous. It is then verified through the numerical experiments that as $n \to \infty$, the detection performance is around 2dB away from the conventional EMR. In addition to formulate the EMR in terms of one-bit samples, the major contributions of this work are two-folded: 1) Calculation of the theoretical decision threshold for the one-bit EMR statistic, which converges very well as $n \to \infty$. 2) Evaluation of the hardware consumption of the reformulated detector that demonstrate the superiority of the one-bit quantization scheme.

\section{EMR Detector with One-bit Sampling}
\subsection{Preliminary on EMR Detector}
Consider a CR network where the SU senses $d$ PUs with $m$ antenna. Denoted by $\mathcal{H}_0: {\boldsymbol{x}\left(t\right)} = {\boldsymbol{\omega}\left(t\right)}$ being the complex data vector in the absence of PU, and $\mathcal{H}_1: {\boldsymbol{x}\left(t\right)} = \boldsymbol{Hs}\left(t\right) + {\boldsymbol{\omega}\left(t\right)}$ being its general alternative, where ${\boldsymbol{s}}\left(t\right) \in  {\mathbb{C}^{d \times 1}}$ and ${\boldsymbol{H}} \in {\mathbb{C}^{m \times d}}$ stand for the unknown signal and channel coefficients, respectively. Moreover, ${\boldsymbol{\omega}\left(t\right)}$ is assumed to be the $i.i.d$ zero mean circular Gaussian vectors with covariance matrix $\tau{{\boldsymbol{I}}_m}$, wherein $\tau$ stands for the unknown noise power. Additionally, it is supposed that the signal vector follows an $i.i.d.$ complex Gaussian distribution with zero mean and covariance matrix $\boldsymbol{R}_{ss}$. Under this circumstance, only the second order statistical is needed to characterize both hypotheses. As $n \to \infty$ and $m \to \infty, m/n \to c$ wherein $c\in \left( 0,1\right)$, \cite{RN1} provided that the formulation of the second order EMR detector based on the Frobenius norm and matrix trace of the SCM, which is given by 
\begin{equation*}
	\label{eq_emr_inf}
	\xi^{\left( 2 \right)} _{\rm{EMR, \infty}} = \frac{\frac{1}{m}{{\left\| \boldsymbol{\Phi} \right\|_F^2}}}{\left(\frac{1}{m} \text{tr} \left(\boldsymbol{\Phi}\right)\right)^2}, 
\end{equation*}
where $ \left\| \cdot \right\|$ and $\text{tr} \left( \cdot \right)$ calculates the Frobenius and trace of square matrix. $\boldsymbol{\Phi} = \left(1 / n\right) \sum_{t = 1}^{n} \boldsymbol{x}\left(t\right) \boldsymbol{x}^H\left(t\right)$ is the SCM of the $\infty$-bit samples (or high resolution quantization) wherein $\left( \cdot \right)^H$ returns the conjugate transpose of the matrix. The subscript $\infty$ signifies the data is $\infty$-bit quantized and the superscript indicates it calculates the second EMR. For simplicity, in the following context, we discard such notations for one-bit samples. Accordingly, we can declare the existence of the PU by judging if
\begin{equation*}
	\xi^{\left( 2 \right)} _{\rm{EMR, \infty}} > \eta^{\left( 2 \right)} _{\rm{EMR, \infty}},
\end{equation*}
where $\eta_{\rm{EMR}}^{\left(2\right),\infty}$ is the corresponding theoretical detection threshold in the generalized asymptomatic regime, which is calculated by
\begin{equation}
	\label{eq_the_inf}
	\eta_{\rm{EMR}}^{\left(2\right),\infty} = 1+c+\frac{\sqrt{2}Q^{-1}\left(1 - \epsilon\right)}{n},
\end{equation}
with $Q\left(\cdot\right)$ being the cdf of the stand normal distribution.

\subsection{Problem Formulation}
On the contrary, one-bit quantization of the $\infty$-bit samples ${\boldsymbol{x}\left(t\right)}$ can be achieve by respectively preserving only the sign of the real and imagine parts in an element-wise manner, and then stack each part into a column. This data acquisition scheme can be expressed by 
\begin{equation}
	\label{eq_onebit_quantize}
	\boldsymbol{z} \left(t\right) = \text{sgn} \left[
	\begin{aligned}
		& {\boldsymbol{x}^R}\left(t\right)  \\
		& {\boldsymbol{x}^I}\left(t\right)
	\end{aligned}
	\right] \in \mathbb{C}^{2m \times 1},
\end{equation} 
where $\text{sgn}\left( \cdot \right)$ preserved the sign of the matrix by element-wised mean. The subscript ``R" and ``I" stands for the real part and the imaginary part of the received data, respectively. 

Similarly to the $\infty$-bit sampling, the SCM for the one-bit counterpart can be thereby calculated as
\begin{equation}
	\boldsymbol{S} = \sum_{t = 1}^{n} { \boldsymbol{z} \left( t \right) \boldsymbol{z}^T \left( t \right)}.
\end{equation}
where ${\left(\cdot\right)^T}$ calculates the transpose of the matrix. It can be easily observed that the diagonal entries of the one-bit SCM are ones under both hypotheses. Under the null hypothesis, in particular, the noise power has been eliminated through this one-bit sampling. Furthermore, note the symmetric property of the SCM, the one-bit EMR of second order can be heuristically expressed by
\begin{equation}
	\label{eq_EMR_onebit}
	{\xi _{\rm{EMR}}} = 1 + \frac{1}{m}\sum\limits_{i = 1}^{2m - 1} {\sum\limits_{j = i + 1}^{2m} {{{\left( {{\boldsymbol{S}}_{i,j}} \right)}^2}} }
\end{equation}
The hypothesis $\mathcal{H}_0$ is then rejected when the test statistic greater than a given threshold $\eta_{\rm{EMR}}$, i.e.
\begin{equation}
	\xi_{\rm{EMR}} \mathop  \gtrless \limits_{{{\cal H}_0}}^{{{\cal H}_1}}  \eta_{\rm{EMR}}.
\end{equation}

\subsection{Theoretical Approximated Detection Threshold}
To assure the CFAR property, the detection threshold is required to declare the existence of the PU. This threshold can be determined by developing the distribution of the test statistic under $\mathcal{H}_0$ and taking the inverse of CDF with a desired false alarm rate ${P}_{fa}$. However, when this extreme quantization scheme is applied, only the sign of the received data remains available. Hence, the sampled data will no longer obey the joint multivariate Gaussian distribution, but become distributed according to the multivariate orthant probability, as discussed in \cite{RN2}. However, the orthant probability remains to be an open problem for dimensions $m \geq 3$. Therefore, it is non-trivial to calculate the threshold either analytically or numerically. 

To deal with this issue, we resort to analyze the test statistic asymptotically as $n \to \infty$. This is practical since we can easily achieve large sample sizes with one-bit ADCs \cite{RN4}. With this distribution, an approximated threshold can be calculated to guarantee the CFAR property. 

To begin with, it is easily observed that the $\left(i,j\right)$th entry of the SCM can be expressed as
\begin{equation}
	\boldsymbol{S}_{i,j} = \frac{1}{n} \sum_{t = 1}^{n} {{\boldsymbol{z}_i}\left(t\right) {\boldsymbol{z}_j} \left(t\right) } =  \frac{1}{n} \sum_{t = 1}^{n} {\rho_{i,j}\left( t \right)}.
\end{equation}
Additionally, denoting
\begin{equation}
	\label{eq_pmf_gamma}
	\begin{aligned}
		\bar{P}_{i,j} &\triangleq \Pr \left\{ \rho_{i,j}\left(t\right)= 1 \right\} \\
		& = \Pr \left\{ {\boldsymbol{z}_i}\left(t\right) = 1, {\boldsymbol{z}_j}\left(t\right) = 1\right\} \\
		& + \Pr \left\{ {\boldsymbol{z}_i}\left(t\right) = -1, {\boldsymbol{z}_j}\left(t\right) = -1\right\},		
	\end{aligned}	
\end{equation}
we have $\Pr \left\{ \rho_{i,j}\left( t \right) = -1 \right\}  = 1 - \bar{P}_{i,j} $, which indicates that $\rho_{i,j}\left( t \right)$ is $i.i.d.$ bivariate distributed. In the light of CLT, as $n \to \infty$, the following corollary is straightforwardly obtained.\\
\emph{Corollary} 1. For sufficiently large sample size $n$, the upper-triangle entries of the noise-only SCM are approximately Gaussian distributed as $\boldsymbol{S}_{i,j} \overset{a.s.}{\sim } \mathcal{N}\left(\bar{\mu},\bar{{\sigma}}^2\right)$, where $\bar{\mu} = 2\bar{P}_{i,j}-1$ and $\bar{{\sigma}}^2 = 4\bar{P}_{i,j}\left(1-\bar{P}_{i,j}\right)/n$.

However, it requires to calculate the 2-dimensional orthant probability to determine (\ref{eq_pmf_gamma}). Fortunately, recall that the spatially white assumption under the null hypothesis, and further taking into account that the real part $\boldsymbol{x}^R\left(t\right)$ and the imaginary part $\boldsymbol{x}^I\left(t\right)$ are mutual independent under $\mathcal{H}_0$, we can calculate this 2-dimensional orthant probability by multiplying two 1-dimensional orthant probability, namely,
\begin{subequations}
	\begin{equation}
		\begin{aligned}
			& \Pr \left\{ {\boldsymbol{z}_i}\left(t\right) = 1, {\boldsymbol{z}_j}\left(t\right) = 1\right\} \\
			& = \Pr \left\{ {\boldsymbol{z}_i}\left(t\right) = 1 \right\} \cdot \Pr \left\{ {\boldsymbol{z}_j}\left(t\right) = 1\right\},
		\end{aligned}
	\end{equation}
	\begin{equation}
		\begin{aligned}
			& \Pr \left\{ {\boldsymbol{z}_i}\left(t\right) = 1, {\boldsymbol{z}_j}\left(t\right) = -1\right\} \\
			& = \Pr \left\{ {\boldsymbol{z}_i}\left(t\right) = 1 \right\} \cdot \Pr \left\{ {\boldsymbol{z}_j}\left(t\right) = -1\right\},
		\end{aligned}
	\end{equation}
	\begin{equation}
		\begin{aligned}
			& \Pr \left\{ {\boldsymbol{z}_i}\left(t\right) = -1, {\boldsymbol{z}_j}\left(t\right) = 1\right\} \\
			& = \Pr \left\{ {\boldsymbol{z}_i}\left(t\right) = -1 \right\} \cdot \Pr \left\{ {\boldsymbol{z}_j}\left(t\right) = 1\right\},
		\end{aligned}
	\end{equation}
	\begin{equation}
		\begin{aligned}
			& \Pr \left\{ {\boldsymbol{z}_i}\left(t\right) = -1, {\boldsymbol{z}_j}\left(t\right) = -1\right\} \\
			& = \Pr \left\{ {\boldsymbol{z}_i}\left(t\right) = -1 \right\} \cdot \Pr \left\{ {\boldsymbol{z}_j}\left(t\right) = -1\right\},
		\end{aligned}
	\end{equation}
\end{subequations}

Meanwhile, for all zero-mean $i.i.d.$ distributed $\boldsymbol{x}_k\left(t\right)$, we have the corresponding probability mass function (pmf) for the real part given by
\begin{subequations}
	\label{eq_pmf_z}
	\begin{equation}
		\Pr \left\{ {\left. {{\boldsymbol{z}}_k = 1} \right|{{\mathcal H}_0}} \right\}= \int_{0}^{ + \infty } {f_{\boldsymbol{X}_k}\left(\left. \boldsymbol{x}_k^{R} \right|{{\mathcal H}_0} \right)}d\boldsymbol{x}_k^{R}  = \frac{1}{2},
	\end{equation}
	\begin{equation}
		\Pr \left\{ {\left. {{\boldsymbol{z}}_k = -1} \right|{{\mathcal H}_0}} \right\} = 1-\Pr \left\{ {\left. {{\boldsymbol{z}_k} = 1} \right|{{\mathcal H}_0}} \right\} = \frac{1}{2},
	\end{equation}
\end{subequations}
for all integer $k \in \left[1,m\right]$, where we discard the time index for the time being. And a similar result can be drawn for the imaginary part with integers $k \in \left[m+1, 2m\right]$. Plugging (\ref{eq_pmf_z}) into (\ref{eq_pmf_gamma}), it is thereby shown $\bar{P}_{i,j} = 1/2$. With some simple manipulations, we have $\sqrt{n}\boldsymbol{S}_{i,j} \overset{a.s.}{\sim } \mathcal{N}\left(0,1\right)$ as $n \to \infty$.

Corollary 1 is built to provide an entry-wise distribution of the one-bit SCM for large $n$. However, to illustrate the statistical property of the $\xi_{\rm{EMR}}$, we need the distribution of the summation of the squares of the upper-triangle entries, thus the inter-entry property needs to be addressed. This can be achieved by using the independence between the upper-triangle entries and the properties of the $\chi^2$ distribution. The result is summarized in the following proposition.\\
\emph{Proposition} 1. Reformulate the upper-triangle entries $\boldsymbol{S}_{i,j}$ for $1 \leq i<j \leq \left(2m\right)$ into a column vector
\begin{equation*}
	\boldsymbol{r} = \left[ \boldsymbol{S}_{1,2}, \boldsymbol{S}_{1,3}, \boldsymbol{S}_{2,3},\cdots, \boldsymbol{S}_{1,2m}, \cdots, \boldsymbol{S}_{2m-1,2m} \right]^T.
\end{equation*} 
Under $\mathcal{H}_0$, the covariance matrix $ \boldsymbol{C}_{\boldsymbol{r}} = \mathbb{E}\left\{ \boldsymbol{r} \boldsymbol{r}^T \right\}$ is diagonal.\\
\emph{Proof}: See Appendix.\qed

Proposition 1 indicates that the test statistic $\xi_{\rm{EMR}}$ can be regarded as the summation of $m\left(2m-1\right)$ squared \emph{i.i.d.} zero-mean Gaussian elements whose variance equals to $1/\sqrt{n}$. To be more concrete, it is straightforward provided that $mn \left( \xi_{\rm{EMR}}-1 \right) \sim \chi^2_q$, where $q = m\left(2m-1\right)$ is the degree of freedom of the Chi-square distribution.
 
The false alarm probability for the one-bit EMR is then calculated as
\begin{equation}
	\begin{aligned}
		P_{fa} & = \Pr \left\{ \left. \xi_{\rm{EMR}} >  \eta_{\rm{EMR}} \right| \mathcal{H}_0  \right\} \\
		& \overset{a.s.}{\simeq } F_{\chi^2_q} \left(  {mn\left( \eta_{\rm{EMR}} -1 \right)}\right)
	\end{aligned}
\end{equation}
where $F_{\chi^2_q} \left(\cdot \right)$ stands for the cdf of Chi-square distribution with degree of freedom $q$. For a desired false-alarm rate $\epsilon$, the corresponding threshold should be chosen as
\begin{equation}
	\label{eq_the_1}
	\eta_{\rm{EMR}} = 1 + \frac{F_{\chi^2_q}^{-1}\left( 1 - \epsilon\right)}{mn}.
\end{equation}

Meanwhile, for large enough $q$, $mn\left(\xi_{\rm{EMR}}-1\right)$ can also be regarded as the summation of $q$ \emph{i.i.d.} $\chi^2_1$ elements. Therefore, its distribution can also be approximated by exploiting the CLT, arriving at 
\begin{equation}
	\label{eq_pdf_emr_asy}
	\frac{nm\left( \xi_{\rm{EMR}}-1 \right) - q}{\sqrt{2q}} \sim \mathcal{N}\left(0,1\right).
\end{equation}
Therefore we can determine the threshold by
\begin{equation}
	\label{eq_the_1_asy}
	\tilde{\eta}_{\rm{EMR}} = 1 + \frac{\sqrt{2q}} {mn}\left( Q^{-1}\left(1 - \epsilon\right)+ \sqrt{\frac{q}{2}}\right).
\end{equation}

Whereas, under the alternative hypothesis, the calculation of $\bar{P}_{i,j}$ requires us to calculate the 2-dimensional orthant probability which is not equal for different pairs of $\left(i,j\right)$. Moreover, deriving the distribution of the test statistic under $\mathcal{H}_1$ relies on the high dimensional orthant probability, which is an open problem. Therefore, we resort to the numerical evaluation in the following section.

\section{Numerical Experiment}
In this section, we report the numerical result of the reformulated one-bit EMR detector comparing with $\infty$-bit EMR detector via Monte Carlo simulations. We first investigate the accuracy of the threshold to validate our theoretical analysis under null hypothesis, which includes both one-bit and $\infty$-bit quantization schemes. Then, we present the detection performance as the function of single to noise power ratio (SNR). Finally, we evaluate the computational complexity in terms of hardware cost for fixed detection probability. Additionally, the false alarm rate is set to be $P_{fa} = 1 \times 10^{-3}$, unless otherwise stated.  
\begin{figure}[!t]
	\centering
	\includegraphics[width=3.4in]{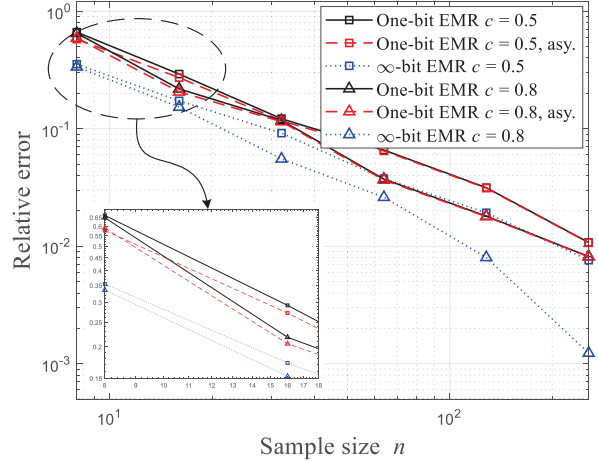}
	\caption{Relative error of the asymptotic/empirical threshold versus sample size for $P_{fa} = 1 \times 10^{-3}$.}
	\label{fig_1}
\end{figure}
Fig.\ref{fig_1} plots the relative errors as the function of the sample size $n$. This error is defined as $\left| \eta_{\text{the}} - {\eta}_{\text{emp}} \right|/ {\eta}_{\text{emp}}$, where $ {\eta}_{\text{emp}}$ is the empirical threshold obtained using $10^3/P_{fa}$ independent trails and $\eta_{\text{the}}$ is the related theoretical threshold. Here we set $c = 0.5$ and $c = 0.8$, refereed to as the generalized asymptotic regime as $n,m \to \infty$ and $m/n \to c$. It is observed that, neither $\eta_{\rm{EMR}}$ nor $\tilde{\eta}_{\rm{EMR}}$ is accurate when $n$ is small. This is because that they are determined by means of CLT, which is valid \emph{only} when the sample size is sufficiently large. On the contrary, (\ref{eq_the_inf}) is determined under random matrix theory, thus providing a good approximation for even small $m$ and $n$. Fortunately, as the sample size $n$ is greater than $2^6$, the relevant error of $\eta_{\rm{EMR}}$ and $\tilde{\eta}_{\rm{EMR}}$ drop below $0.1$, and this error is narrowed as the sample size increases. This implies that the asymptotic distribution of the One-bit EMR statistic is also capable of providing an accurate approximation for sufficiently large $n$. This simulation result is also in accordance with our approximation in (\ref{eq_pdf_emr_asy}).
\begin{figure}[!t]
	\centering
	\includegraphics[width=3.4in]{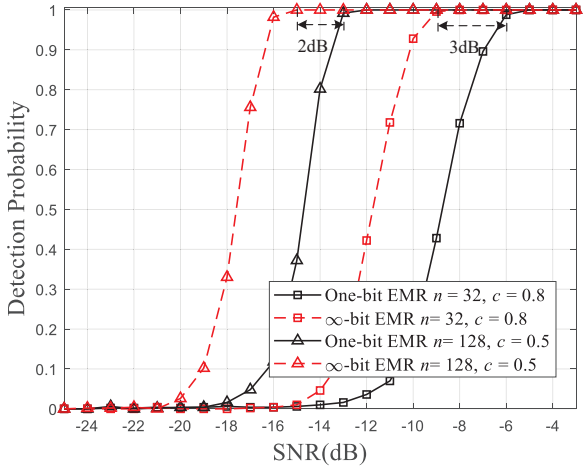}
	\caption{Detection performance versus SNR for determined channel coefficients.}
	\label{fig_2}
\end{figure}
\begin{figure}[!t]
	\centering
	\includegraphics[width=3.4in]{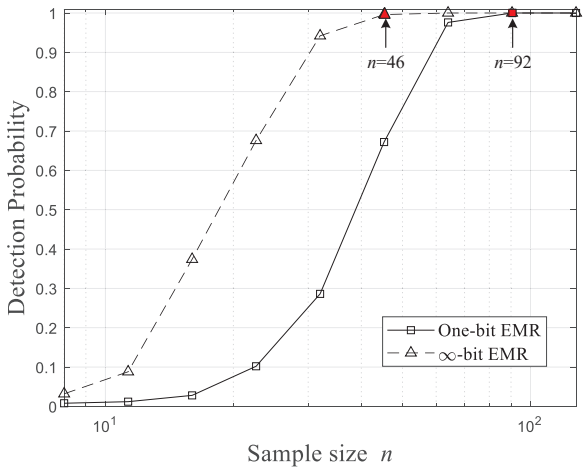}
	\caption{Detection performance versus sample size for $c = 0.5$ and SNR$=-6$dB.}
	\label{fig_3}
\end{figure}

In Fig.\ref{fig_2}, we present the simulation results to evaluate the detection probability. We assume there is only one single-antenna PU located at $-\pi/3$. The SNR is therefore defined by $\sigma ^2_{s}/\tau$. And we consider a SU equipped with an uniformly linear array, the interval between adjacent antenna of whom is of half wavelength. It comes out a $2$dB performance gap is observed when sample size $n = 128, c = 0.5$, which is increased to 3dB when $n = 32, c = 0.8$. This $2$dB gap between one-bit and $\infty$-bit samples have been addressed analytically in \cite{RN6}. Unfortunately, to our best knowledge, the statistic property of $\xi_{\rm{EMR}}$ under $\mathcal{H}_1$ cannot be formulated due to the missing of a closed form expressing of the high dimensional orthant probability. Although the one-bit EMR detector may suffer a performance loss, this gap can be remedied by improving the sample size, which is easily implemented with one-bit ADC. Fig.\ref{fig_3} shows the detection probability with respect to the sample size $n$, where $c = 0.5$ and SNR$=-6$dB are selected for illustration purpose. Under this situation, it requires twice sample size to for the one-bit EMR to achieve the same performance as its $\infty$-bit counterpart. However, the one-bit quantization scheme will obviously reduce the hardware burden in terms of transistor required. This is also the superiority of the one-bit spectrum sensing scheme. 
\begin{table}[!t]
	\caption{Computation Complexity One-bit EMR v.s. $\infty$-bit EMR}
	\label{tab_1}
	\centering
	\begin{tabular}{|p{55pt}|c|c|}
		\hline
		\quad                                 & $\infty$-bit EMR                       & One bit EMR\\
		\hline
		Number of Flops Required \cite{RN1}   & {$m^2\left( n+1 \right)$}              &  $4m^2\left( n+1 \right)$ \\
		\hline
		Number of Transistors Cost            & $1506 \times m^2\left( n+1 \right)$    & $8 \times m^2\left( n+1 \right)$\\
		\hline
	\end{tabular}
\end{table}

To further illustrate this superiority, we evaluate the computational complexity as well as number of transistors costed. However, we use $8$-bit quantization data instead, since it is not realistic to evaluate the $\infty$-bit quantization scheme. As the wireless communicated signals are usually in-phase and quadrate-phase (I/Q) sampled, the multiplications and additions are both performed separately with its real and imaginary parts. According to \cite{RN5}, an 8-bit multiplier costs $748$ transistors, and an 8-bit full adder requires $10$ transistors. So that each 8-bit flops requires $1506$ transistors. Meanwhile, one-bit multiplier can be simplified as an XNOR logical gate, which is constructed by two transistors. Recall the one-bit data formulation in (\ref{eq_onebit_quantize}), where the real and the imaginary part have already been stacked in column-wise, it costs only $2$ transistors for each one-bit flops. Hence, the numbers of transistors required in computing the EMR are listed in Table.\ref{tab_1}. Taking the hardware consumption in Fig.\ref{fig_3} for example, the number of transistors required for 8-bit EMR ($\approx 3.7 \times 10^7$) is approximately $21$ times greater than one-bit EMR ($\approx 1.8 \times 10^6$). In other word, the rapidly sampling one-bit quantization scheme can compensate the performance loss comparing with the slowly sampled data with high precision, whereas the hardware is significantly reduced.

\section{Conclusion}
In this letter, we reformulated the one-bit EMR which inherent the representation of second-order $\infty$-bit EMR. We also derived the asymptotic decision threshold for sufficient large sample size, using CLT. Numerical simulations also illustrate that the analytical threshold is accurate in the general asymptotic regime. Although the detection performance is inferior to its $\infty$-bit counterpart, one-bit ADC allows us to collect data more rapidly, thus making up this performance gap. Interestingly, the one-bit quantization scheme can find its superiority through remarkably reducing the hardware burden, which makes it suitable for computing power starving situation. On the other hand, however, the closed form of detection probability has not yet been addressed. In fact, this distribution is of  theoretical interest, because we can evaluate performance loss more accurately, so that we can select a reasonable sample size accordingly. This will be the subject of our further investigation.

\appendix[Proof of the Proposition 1]
\label{sec_app_A}
By definition, the $\left(p ,l\right)$th entry of $\boldsymbol{C}_{\boldsymbol{r}}$ is calculated as 
\begin{equation}
	\label{eq_cov_r}
	\boldsymbol{C}_{\boldsymbol{r},p,l} = \frac{1}{n^2}\mathbb{E} \left\{ \sum_{t_1 = 1}^{n} \boldsymbol{z}_{i_1} \left(t_1\right)\boldsymbol{z}_{j_1}\left(t_1\right)  \sum_{t_2 = 1}^{n}  \boldsymbol{z}_{i_2} \left(t_2\right) \boldsymbol{z}_{j_2}\left(t_2\right)  \right\},
\end{equation}
where $p, l \in \left[ 1, {m\left(2m-1\right)} \right]$ and $p = i_1 + \frac{\left(j_1-1\right)\left(j_1-2\right)}{2}$ and $l = i_2 + \frac{\left(j_2-1\right)\left(j_2-2\right)}{2}$ for $1 \leq i_1<j_1 \leq 2m, 1 \leq i_2<j_2 \leq 2m$.

We firstly investigate the non-diagonal entries, wherein $p \neq l$. Since $\boldsymbol{z}_k\left( t \right)$ is independent with respect to time index $t$, and $\mathbb{E}\left\{ \boldsymbol{S}_{i,j} \right\} = 0$ for large $n$, the elements in (\ref{eq_cov_r}) with different time index are equals to zero. This simplifies (\ref{eq_cov_r}) into
\begin{equation}
	\label{eq_cov_r_n}
	\boldsymbol{C}_{\boldsymbol{r},p,l} = \frac{1}{n^2}\mathbb{E} \left\{ \sum_{t = 1}^{n} \boldsymbol{z}_{i_1} \left(t\right)\boldsymbol{z}_{j_1}\left(t\right) \boldsymbol{z}_{i_2} \left(t\right) \boldsymbol{z}_{j_2}\left(t\right)  \right\}.
\end{equation}
Since $i_1 = i_2$ and $j_1 = j_2$ if and only if $p = l$, there might exist at least two element with different subscript in calculating (\ref{eq_cov_r_n}). Under $\mathcal{H}_0$, however, recall that $\mathbb{E} \left\{ \boldsymbol{z}\left( t \right) \right\} = \boldsymbol{0}_{2m \times 1}$ and $\boldsymbol{z}_k \left(t\right)$'s are independent for $k \in \left[1, 2m\right]$. It is then provided that $\boldsymbol{C}_{\boldsymbol{r}}\left(p,l\right) = 0$ for all $p \neq l$.

Meanwhile, for diagonal entries, i.e. $p = l$, it is manipulated by $\boldsymbol{C}_{\boldsymbol{r},p,p} = \mathbb{E} \left\{ \boldsymbol{S}_{i,j}^2 \right\} = \bar{{\sigma}}^2$, for sufficiently large $n$,

\ifCLASSOPTIONcaptionsoff
  \newpage
\fi

\bibliography{ref}
\bibliographystyle{IEEEtran}

\end{document}